\documentclass[aps,twocolumn,pre,showpacs,floatfix]{revtex4}
\usepackage{graphics,amssymb,amsmath}

\begin{document}

\title{Optimal Size of a Complex Network}
\author{ H. \surname{Hong}}
\email{hhong@kias.re.kr}
\affiliation{Korea Institute for Advanced Study, Seoul 130-012, Korea}
\author{Beom Jun \surname{Kim}}
%\email{kim@tp.umu.se}
\affiliation{Department of Molecular Science
     and Technology, Ajou University, Suwon 442-749, Korea}
\author{M.Y. \surname{Choi}}
%\email{mychoi@phya.snu.ac.kr}
\altaffiliation{Also at Korea Institute for Advanced Study, Seoul 130-012, Korea.}
\affiliation{Department of Physics, Seoul National University,
Seoul 151-747, Korea}

\begin{abstract}
We investigate the response behavior of an Ising system, driven by an
oscillating field, on a small-world network, with particular attention to the
effects of the system size.  The responses of the magnetization to the driving
field are probed by means of Monte Carlo dynamic simulations with the rewiring
probability varied.  It is found that at low and high temperatures the
occupancy ratio, measuring how many spins follow the driving field, behaves
monotonically with the system size.  At intermediate temperatures, on the other
hand, the occupancy ratio first grows and then reduces as the size is
increased, displaying a resonance-like peak at a finite value of the system
size.  In all cases, further increase of the size eventually leads to
saturation to finite values; the size at which saturation emerges is observed
to depend on the temperature, similarly to the correlation length of the
system. 
%%%%%%%%%%%%%%%%%%%%%%%%%%%%%%%%%%%
%We investigate the effects of the system size on collective response in 
%the field-driven Ising system on a small-world network. 
%The response of the magnetization to an oscillating magnetic field is 
%probed by means of Monte Carlo dynamic simulations, varying the 
%rewiring probability. 
%It is found that the system exhibits the maximized stochastic flip 
%following the external field at a certain value of the system size for 
%any finite value of the rewiring probability, and that size is 
%observed to be increased as the temperature is increased, which is similar
%to the behavior of the correlation length of the system.
\end{abstract}

\pacs{89.75.Hc, 75.10.Hk}
%89.75.-k: Complex systems
%89.75.Fb: Structures and organization in complex systems
%75.10.Hk: Classical spin models

\maketitle

%\section{Introduction}
It has been known that under appropriate circumstances, the presence of noise 
in a driven system can enhance rather than suppress the output of the system.  
Such attractive phenomena, called stochastic resonance (SR), have been widely 
investigated in various systems of many practical 
applications~\cite{ref:SR:general,ref:SR:practical}. 
Recently, those SR phenomena have also been examined in the context of the system size 
resonance: Stochastic flips of the mean field are observed to depend on 
the system size, leading the linear response of the system to reach a maximum at 
a certain system size~\cite{ref:SSR}. 
Those works considered two systems of fully coupled noisy
oscillators and one two-dimensional Ising system with nearest neighbor interactions. 
Namely, the underlying connection topology of dynamic variables was assumed 
to be regular, with either local or global connections. 
Meanwhile, recent studies of neuronal networks, computer networks, 
biochemical networks, and even social networks, have revealed that various real 
systems in nature possess quite complex structures, which can be 
described neither by regular networks nor by completely random 
networks~\cite{ref:network}.  Accordingly, it is desirable to study
effects of the system size on collective responses 
in the systems with the connection topology of complex networks, 
which can be made more realistic. 
In particular, the interplay between the system size and noise may
be relevant in various biological systems
such as neural networks and other cell networks, which consist of 
finite numbers of elements. 
For example, in the study of stochastic resonance in biological systems,
optimal sizes of calcium ion channel clusters have been examined.  Observed is that 
the clustering of the release channels in small clusters increases the sensitivity of 
the calcium response~\cite{ref:Jung}.  This suggests a possible realization of the system,
providing motivation for the investigation of the size resonance in the 
complex network structure. 

In this paper we consider an Ising model on 
a complex network, specifically, on the Watts and Strogatz (WS) type 
small-world network~\cite{ref:WS}. 
It is well known that the WS network is characterized by a small characteristic path 
length $\ell \sim \ln N$, where $N$ is the number of nodes constituting the network, 
and a large clustering coefficient.  Both are commonly observed 
properties of real networks in nature. 
The WS network in this paper is constructed following Ref.~\cite{ref:WS}:
A regular one-dimensional (1D) network of $N$ nodes is first constructed with 
local connections of range $k$, under periodic boundary conditions.
At this stage each node on the network has $2k$ nearest neighbors. 
Next, each local link is visited once and, with the rewiring probability $P$,
removed and reconnected to a randomly chosen node.  After a
whole sweep of the entire network, the average number of shortcuts 
in the network of size $N$ is given by $NPk$.  Accordingly, the rewiring 
probability $P$ may be regarded as the fraction of the average number of shortcuts 
over the total number of connections $Nk$. 
In this paper, the local interaction range $k$ is set equal to 
two for convenience;
longer ranges ($k >2$) are not expected to lead to any qualitative difference.
After the WS network is built as above, an Ising spin is put on every node, 
and an edge (or a link) connecting two nodes is regarded as the coupling
between the two spins at the two nodes.
Finally, we apply an oscillatory field, driving the Ising spins,
and the corresponding responses of the average spin, i.e., the magnetization are 
probed via Monte Carlo dynamic simulations, with attention to the effects of 
the system size. 

The Hamiltonian for the field-driven Ising model on the WS network, 
which is constructed as described above, reads  
\begin{equation}
H = -\frac{1}{2}\sum_{i, j} J_{ij}\sigma_i \sigma_j 
-h(t)\sum_{i}\sigma_i,
\label{eq:Hamiltonian}
\end{equation}
where the ferromagnetic spin-spin interaction strength $J_{ij}$ is given by
\begin{equation} \label{eq:Jij}
J_{ij}  = J_{ji} \equiv \left\{
\begin{array}{ll}
J & \mbox{for $j \in \Lambda_i$,} \\
0 & \mbox{otherwise.}
\end{array}
\right.
\end{equation}
The neighborhood $\Lambda_i$ of node $i$ stands for
the set of nodes connected to $i$ (via either local edges or shortcuts), 
and $\sigma_i \,(=\pm 1)$ represents the Ising spin at node $i$.  
The sinusoidally oscillating magnetic field $h(t)=h_0 \cos\Omega t$ is applied
with the driving amplitude $h_0$ and frequency $\Omega$, while the system is 
assumed to be in contact with an isothermal heat bath at temperature $T$. 
We probe the dynamics of the system described by Eq.~(\ref{eq:Hamiltonian}) 
by means of Monte Carlo (MC) dynamic simulations, employing
the heat bath single spin-flip algorithm~\cite{ref:Newmanbook} and measuring
the time $t$ in units of the MC time step.
For thermalization, we start from sufficiently high temperatures and
lower the temperature $T$ slowly with the decrement $\Delta T = 0.01$ 
(in units of $J/k_B$ with the Boltzmann constant $k_B$). 
The driving amplitude and frequency are taken to be $h_0=0.1$ and 
$\Omega=0.001$.  We have also considered different 
frequencies, for example, $\Omega=0.01$ and $0.1$, and found 
that the resonance-like peak indicating the system size resonance 
behavior tends to diminish at higher frequencies (see below). 
%Due to those reasons, we have chosen the smaller value of the driving 
%frequency such as $\Omega=0.001$. 
%%%%%%%%%%%%%%%%%%%%%%%%%%%%%%%%%%%
%We have also performed MC simulations for the larger driving frequencies 
%such as $\Omega=0.01$ and $0.1$, to find that for larger one the 
%time-dependent magnetization do not tend to oscillate about zero 
%following the magnetic field since the system does not have enough time 
%to switch between the two minima of the free energy during the half of the 
%period of external field.  Therefore, the regime of the smaller driving 
%frequencies is observed to be better one for examining the system 
%size resonance phenomena. 
%%%%%%%%%%%%%%%%%%%%%%%%%%%%%%%%%%%
While simulations are performed at a given temperature, 
the data from the first $10^5$ MC steps are discarded, 
which turns out to be sufficient for stationarity,
and measurements are made for next $10^5$ MC steps.
Networks of various sizes are constructed as described above, and 
averages are taken over 100 different network realizations for each size.
%%%%%%%%%%%%%%%%%%%%%%%%%%%%%%%%%%%%%%%%
%For the undriven ($h_0=0$) Ising model on WS type small-world networks
%it has been addressed that the long-range order exists for any nonzero 
%rewiring probability ($P\neq 0$)~\cite{ref:smallIsing}.
%Meanwhile, the driven ($h_0 \neq 0$) Ising model on WS networks 
%has been studied~\cite{ref:SR_WS_1}, and the double SR peaks which 
%originate from the matching of two time scales has been observed.
%In this paper we consider the same system as in Ref.~\cite{ref:SR_WS_1}, 
%however in the different point of view such as system size resonance behavior.
%%%%%%%%%%%%%%%%%%%%%%%%%%%%%%%%%%%%%%%%

To investigate the collective response of the system, 
we measure the occupancy ratio $R$ which is defined to be the average fraction 
of the spins in the direction of the external field~\cite{ref:ORfirst,ref:BJKim}: 
\begin{equation}
R\equiv \left\langle\frac{\mbox{number of spins in the direction of}~h(t)}
{\mbox{total number of spins}}\right\rangle , 
\label{eq:OR}
\end{equation}
where $\langle\cdots\rangle$ denotes the time average.
In other words, $R$ measures how many spins follow the oscillating 
magnetic field.  It is easy to understand 
that $R$ approaches the value 1/2 in both low- and high-temperature limits
(see, e.g., Ref.~\cite{ref:SR_WS_1,ref:BJKim}) and
becomes increased near the stochastic resonance temperature, reflecting that 
more spins follow the external driving. 
Such SR phenomena have been observed in the system of given size,
and it has been demonstrated that the matching condition of two time scales,
the relaxation time of the system and the inverse frequency of the driving field,
yields the optimal noise strength $T_{SR}$ at which the system displays 
maximum responses~\cite{ref:SR_WS_1,ref:BJKim}. 
Here we consider the system studied in Ref.~\cite{ref:SR_WS_1} from a
different point of view and examine the behavior of the 
occupancy ratio with the system size at various temperatures, 
probing the size resonance.

In the absence of long-range shortcuts ($P=0$),
the network structure reduces to that of the 1D
regular network with only local couplings of range $k$.
Accordingly, when $P=0$, the field-driven Ising model described by 
Eq.~(\ref{eq:Hamiltonian}) as well as the undriven model [$h(t)=0$] does not 
exhibit long-range order at finite temperatures, yielding $T_c=0$. 
Note, however, that even such a 1D system displays
SR behavior at finite temperatures~\cite{ref:Brey}. 
In the presence of long-range shortcuts ($P \neq 0$), on the other hand, 
it has been found that both driven and undriven Ising model
display ferromagnetic order at finite temperatures~\cite{ref:SR_WS_1,ref:smallIsing}; 
furthermore, double SR peaks, which originate from matching of two time scales 
have been observed.

\begin{figure}
\centering{\resizebox*{!}{5.5cm}{\includegraphics{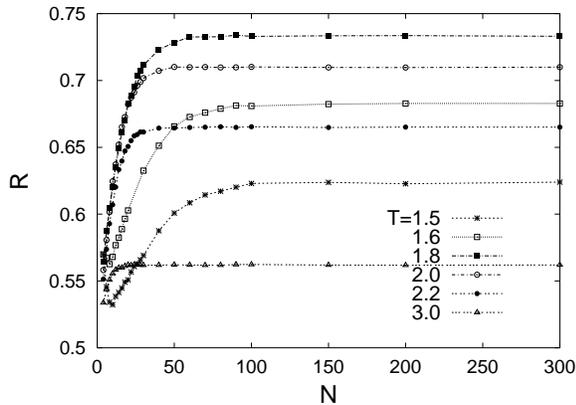}}}
\caption{Behavior of the occupancy ratio $R$ with the system size $N$ at various 
temperatures $T$ in the absence of long-range shortcuts. 
Lines are merely guides to eyes.}
\label{fig:OR_P0.0}
\end{figure}

\begin{figure}
\centering{\resizebox*{!}{5.5cm}{\includegraphics{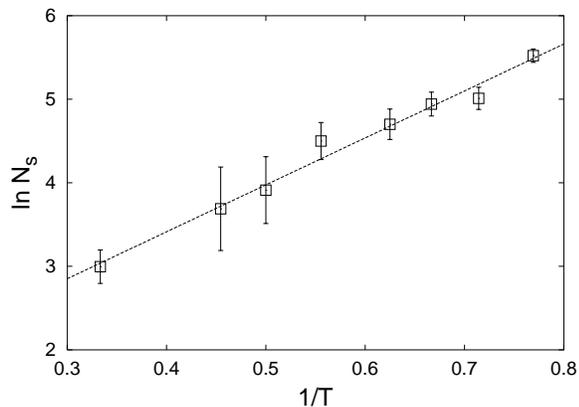}}}
\caption{Behavior of the saturation size $N_s$ with the temperature $T$
in the absence of shortcuts,
exhibiting a linear relation between $\ln N_s$ and $1/T$,
with the slope $c=5.6 \pm 0.7$.
The dotted line, obtained by the least-square fit, represents
$\ln N_s = 5.6/T +1.2$.
}
\label{fig:N0}
\end{figure}

We first consider the case without long-range shortcuts ($P=0$),
i.e., the purely 1D system,
and show in Fig.~\ref{fig:OR_P0.0} the behavior of the occupancy ratio $R$ 
versus the system size $N$ at various temperatures.
It is observed that the occupancy ratio $R$ 
%in the low temperature 
%limit $(T \lesssim 1.6)$ is first decreased for the small size 
%($N\lesssim 10$) and then it 
first increases and eventually saturates to a finite 
value as the system size $N$ is increased. 
The saturation size $N_s$, beyond which the occupancy ratio $R$ 
does not increase any more, reduces as the temperature $T$ is raised. 
Figure~\ref{fig:N0} displays such temperature dependence of the saturation size $N_s$, 
which has been taken as the size giving the occupancy ratio with the 
difference from the stationary value less than $5\times 10^{-4}$. 
One can observe the exponential behavior: $N_s \propto e^{c/T}$ with
$c = 5.6\pm 0.7$, 
which is reminiscent of the behavior of the correlation length. 
In the 1D Ising model the correlation length $\xi$ 
diverges in the low-temperature limit as 
$\xi \sim e^{k(k+1)/T}$~\cite{ref:smallIsing}. 
Here the local interaction range $k$ has been chosen to be two $(k=2)$,
leading to the behavior $\xi \sim e^{6/T}$, 
essentially the same as that of the saturation size. 
It is thus concluded that the saturation behavior emerges when the system 
size reaches the correlation length of the system. 

\begin{figure}
\centering{\resizebox*{!}{5.5cm}{\includegraphics{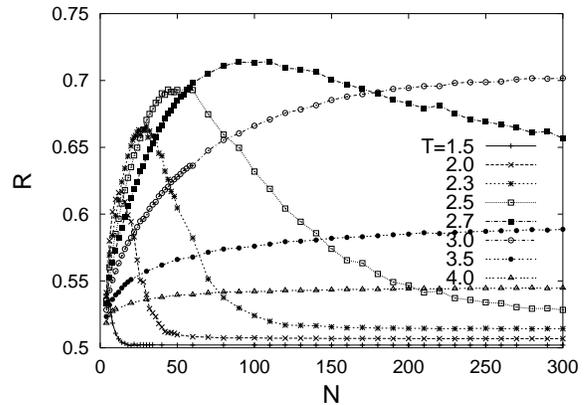}}}
\caption{The occupancy ratio $R$ vs the system size $N$ in the 
presence of the long-range shortcuts ($P=0.5$) at various temperature $T$.
Lines are merely guides to eyes.}
\label{fig:OR_P0.5}
\end{figure}

Meanwhile, in the presence of long-range shortcuts $(P\neq 0)$, 
substantially different behavior has been obtained for the occupancy ratio. 
In Fig.~\ref{fig:OR_P0.5}, the occupancy ratio $R$ of the system with 
the rewiring probability $P=0.5$ is displayed as the system size $N$ 
is varied.
At low temperatures $(T \lesssim 1.5)$, $R$ first decreases
monotonically with the size $N$ and then saturates to the value $0.5$;
at high temperatures ($T\gtrsim 2.9$), on the contrary, $R$ increases
monotonically to the saturation value larger than $0.5$, depending on 
the temperature. 
In contrast to these monotonic behaviors, at intermediate 
temperatures $(1.5\lesssim T \lesssim 2.9)$, the occupancy ratio $R$ 
behaves non-monotonically, exhibiting a maximum at a finite value 
of the system size $N$. 
The height of such a resonance-like peak tends to increase 
as the temperature is raised. 
We have also considered different values of the rewiring probability $P$
as well as of the driving frequency $\Omega$. 
It is found that as $P$ is increased, the range of the temperature, 
in which the size-resonance behavior is displayed, becomes wider 
and that the saturation temperature beyond which 
$R$ shows saturation behavior increases. 
On the other hand, as the driving frequency is increased, the position of the 
resonance peak shifts toward smaller values of the system size, 
thus tending to yield monotonic decrease of $R$. 

\begin{figure}
\centering{\resizebox*{!}{5.5cm}{\includegraphics{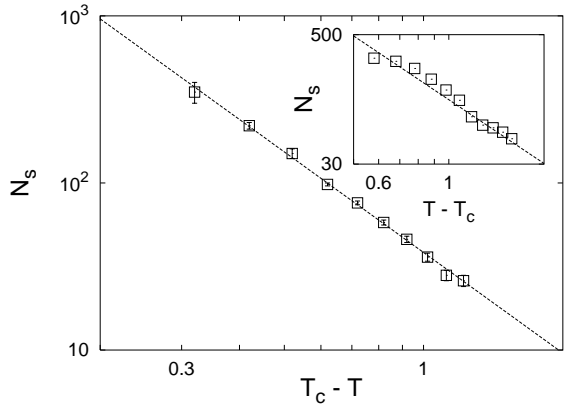}}}
\caption{Behavior of the saturation size $N_s$ with the temperature $T$ 
for $P=0.5$ at low temperatures ($T<T_c$), where the value $T_c =2.7$ obtained from 
the fitting has been used. 
The least-square fit represented by the dashed line corresponds to 
$\log N_s = -2.0 \log (T_c -T) +1.6$, the slope of which gives $\bar\nu = 2.0$. 
Inset: Behavior of $N_s$ at temperatures higher than $T_c$. 
The dashed line, obtained from the least-square fit, is given by 
$\log N_s = -2.0 \log (T-T_c ) +2.1$, where the slope again leads to
$\bar\nu = 2.0$.  The error bars have sizes not larger than the symbol size.
}
\label{fig:Ns_Tupdown}
\end{figure}

\begin{figure}
\centering{\resizebox*{!}{5.5cm}{\includegraphics{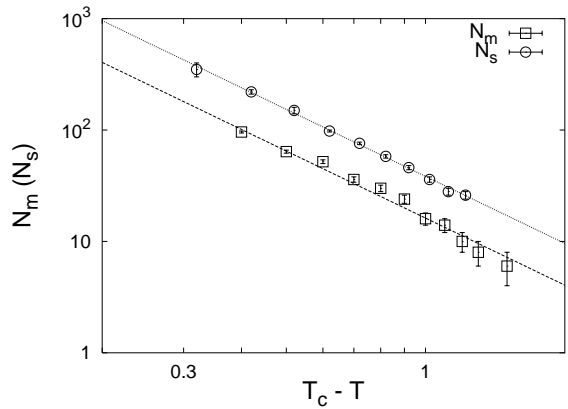}}}
\caption{The resonance size $N_m$ together with the saturation size $N_s$ 
vs the temperature $T$ in the presence of shortcuts ($P=0.5$).  
The values of $T_c$, obtained from the fitting, are given by $3.1$ and $2.7$ 
for $N_m$ and $N_s$, respectively. 
The dotted line represents the corresponding least-square fit of $N_m$, 
which is described by $\log N_m = -2.0 \log (T_c -T) +1.2$. 
Thus both $N_m$ and $N_s$ result in the same value $\bar\nu = 2.0$.
}
\label{fig:Ns_Nm}
\end{figure}

\begin{figure}
\centering{\resizebox*{!}{5.5cm}{\includegraphics{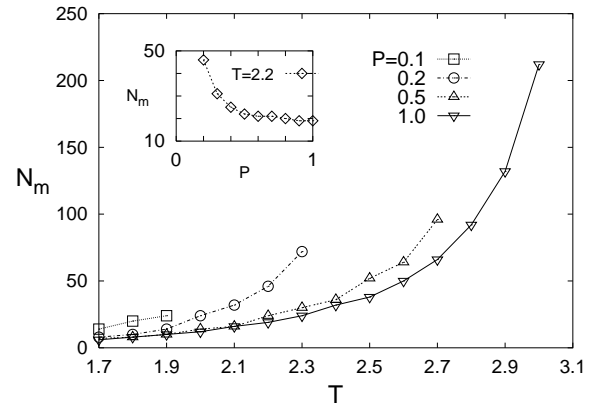}}}
\caption{The resonance size $N_m$ vs the temperature $T$ at 
various values of the rewiring probability $P$.  Inset: Behavior of $N_m$ 
with the rewiring probability $P$ at temperature $T=2.2$.}
\label{fig:Nmax}
\end{figure}

Note that this resonance behavior manifests two kinds of length scale in the 
system with long range shortcuts:
the saturation size $N_s$ and the resonance size $N_m$ 
at which $R$ reaches the maximum. 
To understand the possible relation with the correlation length
even in the presence of long range shortcuts, 
we examine the behaviors of $N_s$ and $N_m$, which are displayed in
Figs.~\ref{fig:Ns_Tupdown} -~\ref{fig:Nmax}.

Figure~\ref{fig:Ns_Tupdown} exhibits the saturation size $N_s$ versus
the temperature $T$ in the system with $P=0.5$.  
For comparison with the correlation volume 
described by $\xi_{V} \sim |T-T_c|^{-\bar\nu}$~\cite{ref:XY},
the data points are plotted in the logarithmic scale, thus
fitted to a linear relation between $\ln N_s$ and $\ln |T-T_c|$ 
with the proportionality constant (slope) $\bar\nu$. 
%with the control parameters $\bar\nu$, $T_c$, and $b$ adjusted.
From this fitting, we obtain $T_c \approx 2.7$, with which
the slope is estimated to be $\bar\nu = 2.0 \pm 0.1$.
Noting that in $d$ dimensions, the correlation volume relates with the correlation 
length $\xi$ via $\xi_V \sim \xi^d$ and the behavior $\xi \sim |T-T_c|^{-\nu}$,
we thus have ${\bar \nu} = d\nu$ in a $d$-dimensional system.
Here it is known that the (effective) dimension $d$ of a mean-field system should
be taken as the upper-critical dimension $d_u$~\cite{ref:Botet}, leading to
$\bar\nu=d_u \nu$.  With $d_u =4$ and $\nu=\nu_{MF} =1/2$ for a mean-field system,
we conclude that the value $\bar\nu=2.0$ indicates a transition of the 
mean-field nature~\cite{ref:XY}.
We also examine the other length scale $N_m$
and show its temperature dependence in Fig.~\ref{fig:Ns_Nm}, 
where for convenience $N_s$, shown already in Fig.~\ref{fig:Ns_Tupdown}, 
is also plotted.  
It is shown that both the two length scales behave similarly with the temperature, 
yielding essentially the same value of the exponent 
$\bar\nu = 2.0 \pm 0.1$ (see the slopes of the two fitted lines). 
Accordingly, both the two length scales $N_s$ and $N_m$ apparently 
measure the correlation length of the system. 
On the other hand, the fitting parameter $T_c$ for $N_m$ turns out to be 
$3.1$, which is somewhat higher than the value $2.7$ obtained from 
fitting of $N_s$.  At this stage it is difficult to discern unambiguously 
the difference; more extensive simulations and careful analysis 
are necessary for confirming the origin as well as the presence of this discrepancy. 

Finally, we consider systems with different rewiring probabilities,
and examine how the rewiring probability $P$ affects the resonance behavior. 
Figure~\ref{fig:Nmax} shows the behavior of the resonance size $N_m$ 
with the temperature at various values of $P$.
It is observed that $N_m$ first increases slowly with the temperature $T$
then very fast as $T$ approaches $T_c$, which again reminds us of the behavior 
of the correlation length. 
Behavior of $N_m$ with the rewiring probability $P$ at given 
temperature $T=2.2$ is displayed in the inset of Fig.~\ref{fig:Nmax}. 
Note the rather fast decrease of $N_m$ for small rewiring probabilities 
($P \lesssim 0.5$) and the saturation behavior for large rewiring probabilities 
($P \gtrsim 0.5$).  
Such saturation behavior has also been reported in the synchronization 
of the system of coupled oscillators on a WS network~\cite{ref:Hong}. 
It is also noteworthy that $N_m$ decreases as shortcuts are
introduced to the system;
the shortcuts tend to decrease the optimal system size 
which corresponds to the maximum collective response at given 
temperature and rewiring probability.

In conclusion, we have investigated the effects of the system size 
on the collective response, measured by the occupancy ratio, 
in the oscillatory field-driven Ising model on WS networks.  
In the purely one-dimensional system without long-range interactions, 
which does not undergo a phase transition at any finite temperature,
the occupancy ratio has been found to display monotonic behavior, 
not exhibiting a resonance peak.
As long-range interactions come into the system, on the other hand, 
system-size resonance, characterized by nonmonotonic behavior, has been observed 
to emerge, thus suggesting a possible relation between the size resonance
and a finite-temperature phase transition. 
The resonance size at which the occupancy ratio reaches the peak
may be regarded as the optimal size of the network, 
in view of the maximum response.
%
%The interval of the temperature where such system size resonance 
%behavior sets in is observed to be increased as the rewiring probability 
%$P$ is increased.
%We note that such system size resonance behavior is not shown in the case of 
%large driving frequency ($\Omega \gtrsim 0.01$). 
%We have performed MC simulations for the larger driving frequencies 
%such as $\Omega=0.01$ and $0.1$, to find that for larger one the 
%time-dependent magnetization do not tend to oscillate about zero value
%following the magnetic field since the system does not have enough time 
%to switch between the two minima of the free energy during the half of the 
%period of external field.  Therefore, the regime of the smaller driving 
%frequencies is observed to be better one for examining the system 
%size resonance phenomena. 
%%%%%%%%%%%%%%%%%%%%%%%%%%%%%%%%%%
It is noteworthy that the optimal size as well as the saturation size 
displays temperature-dependent behavior, which
is essentially the same as that of the correlation length of the system. 
This suggests the interesting possibility of 
estimating the correlation length from the stochastic
resonance behavior at various sizes.  Namely, the
size resonance phenomena may be used as a tool to measure the correlation length.
Note also that at given temperature both length scales, 
the optimal size and the saturation size,
tend to decrease as the amount of long-range interactions is increased.

As a possible application of the system size resonance, we suggest 
biological systems such as the assembly of beta cells which reside 
in the islets of Langerhans in a pancreas~\cite{ref:Lange}. 
It is known that the beta cells form clusters, each with a finite number 
of cells, rather than gathering together as one unit. 
Thus speculated is the possibility that the function of the beta cells may be 
optimized via the mechanism of the system-size resonance. 
In addition, the system size resonance behavior may also be useful for 
understanding the formation of the public opinion~\cite{ref:opinion}.

\acknowledgments
H.H. thanks J. Lee for providing the privilege of using the computing facility Gene.  
B.J.K. was supported in part by the Korea Science and Engineering
Foundation through Grant No. R14-2002-062-01000-0.
M.Y.C. thanks the Korea Institute for Advanced Study for hospitality during his visit, 
where this work was performed, and
acknowledges the partial support from 
%the Korea Science and Engineering Foundation through 
the Swiss-Korean Outstanding Research Efforts Award Program. 
%and the Ministry of Education through the BK21 Program.

\end{document}